\documentclass[conference]{IEEEtran}

\usepackage{cite}
\usepackage{amsmath,amssymb,amsfonts}
\usepackage{algorithmic}
\usepackage{algorithm}
\usepackage{graphicx}
\usepackage{textcomp}
\usepackage{xcolor}
\usepackage{url}
\usepackage{listings}
\usepackage{tabularx}
\usepackage{hyperref}

\begin{document}
\title{CoGen: Creation of Reusable UI Components in Figma via Textual Commands}
\date{}%date stay empty

\author{
\IEEEauthorblockN{ Ishani Kanapathipillai}
\IEEEauthorblockA{ Informatics Institute of Technology \\  Colombo, Sri Lanka \\ University of Westminster \\London, United Kingdom \\ ishani.20200139@iit.ac.lk\\}
\and
\IEEEauthorblockN {Obhasha Priyankara }
\IEEEauthorblockA{Informatics Institute of Technology\\ Colombo, Sri Lanka \\ obhasha@live.com\\}}
\maketitle

\begin{abstract}
The evolution of User Interface design has emphasized the need for efficient, reusable, and editable components to ensure an efficient design process. This research introduces CoGen, a system that uses machine learning techniques to generate reusable UI components directly in Figma, one of the most popular UI design tools. Addressing gaps in current systems, CoGen focuses on creating atomic components such as buttons, labels, and input fields using structured JSON and natural language prompts.
% The project involves two major components: 

The project integrates Figma API data extraction, Seq2Seq models, and fine-tuned T5 transformers for component generation. The key results demonstrate the efficiency of the T5 model in prompt generation, with an accuracy of 98\% and a BLEU score of 0.2668, which ensures the mapping of JSON to descriptive prompts. For JSON creation, CoGen achieves a success rate of up to 100\% in generating simple JSON outputs for specified component types.

\end{abstract}

\section{Introduction}
User Interface (UI) designs play a vital role in bridging human interaction and digital systems~\cite{pratama_effect_2020}. The development of efficient UI designs plays an essential role in enhancing the accessibility, functionality, and the overall appeal of digital products~\cite{jain_sketch2code_2019}. UI design involves following the design thinking process, which includes empathizing, defining, ideating, prototyping and testing~\cite{manandhar_magic_2021}. The important stages of this process range from the creation of wireframes, low fidelity (Lo-Fi), and high fidelity (Hi-Fi) prototypes to more generic stages such as the creation of design systems~\cite{lamine_understanding_2022}.

% \subsection{UI components}
UI components are generally created during the initial stages of the design thinking process in any UI design project~\cite{odushegun_aesthetic_2023, alao_user-centereduser_2022}. The authors in~\cite{bakaev_component-based_2018} state that a component is a structural unit that is an ``abstract element in the structure of the system, solving certain subtasks within the goals of the system, and interacting with the environment through the interface(s)``. 

\begin{figure}
    \centering
    \includegraphics[width=1\linewidth]{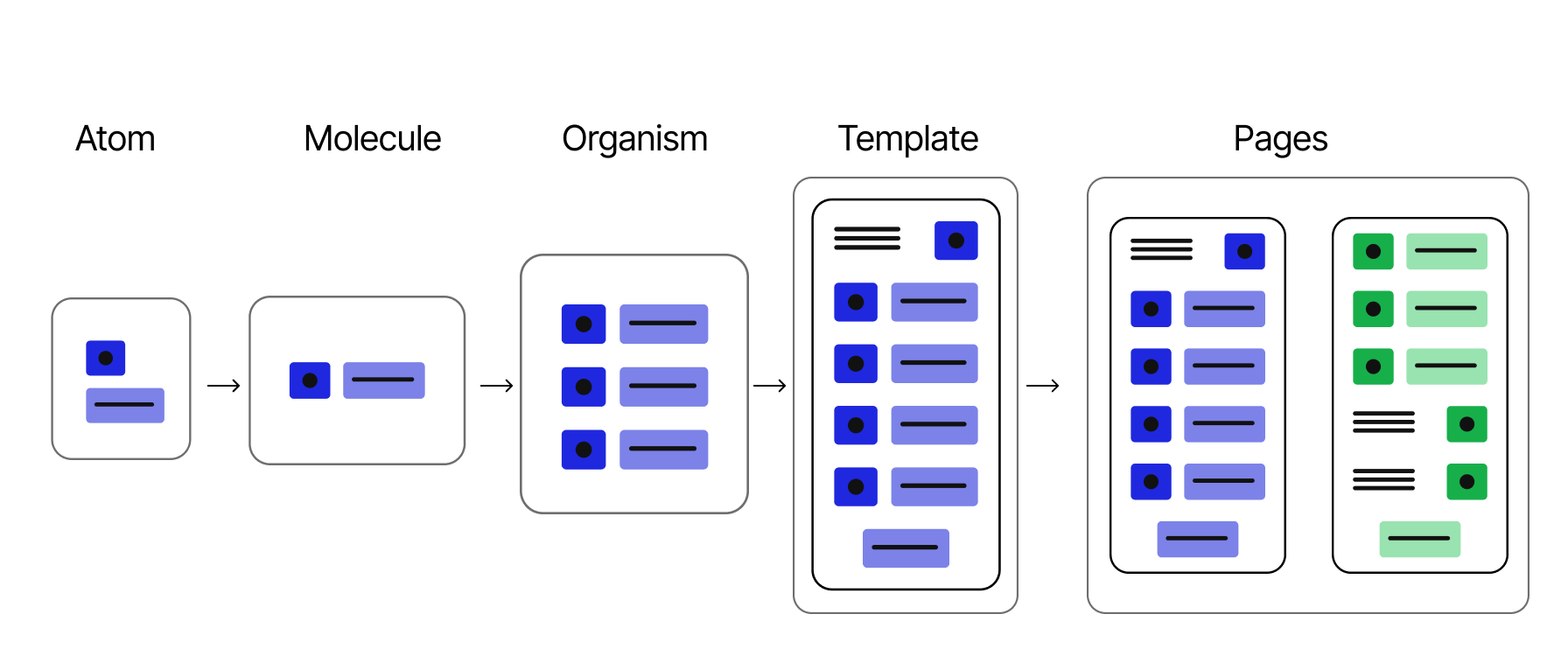}
    \caption{Atomic design system}
    \label{fig:atom}
\end{figure}

UI design systems consist of components that encompass the five stages of the Atomic Design concept~\cite{gustafsson_defining_2021}. This concept consists of \textit{atoms}, \textit{molecules}, \textit{organisms}, \textit{templates}, and \textit{pages}~\cite{tidwell_designing_2019}. %Each stage is the combination of one or more components from the previous stage~\cite{augusdi_development_2021}. 
%
% \subsection{Atoms in Design Systems}
Atoms are the fundamental building blocks that act as the foundation for the next stages in the Atomic Design Concept~\cite{moran_machine_2020}. Atoms can be anything like buttons, labels, or input fields~\cite{tidwell_designing_2019}. These basic components can be combined to create the rest of the atomic design concept~\cite{augusdi_development_2021}, as shown in Fig~\ref{fig:atom}. 
%Most projects in UI design focus on the latter part of the atomic design concept. %

Extensive research has been conducted in the UI design field, with the majority of studies using ML techniques like Generative Adversarial Networks (GAN), Sequence-to-Sequence (Seq2Seq) models to automate various aspects of the design process. However, these projects mostly focus on generating wireframes, Lo-Fi, and Hi-Fi prototypes, and code generation, while overlooking the critical task of creating individual UI components. Despite their role as the building blocks of design systems, the automated generation of reusable and editable UI components is yet to be explored.

Accordingly, the primary research gaps we identified are as follows:

\begin{itemize}
    \item \textit{Unavailability of systems to generate UI components automatically:} Most currently available systems in UI design are geared toward either handling designs associated with textual terms~\cite{lu_ui_2023} or producing designs using linguistic parsers based on non-neural techniques\cite{huang_creating_2021}. Additionally, more emphasis has been placed on UI design generation where authors have developed systems that create whole UIs based on input data. These designs range from wireframes~\cite{gajjar_akin_2021} and Lo-Fi prototypes~\cite{acornley_using_2021} to Hi-Fi prototypes~\cite{zhao_guigan_2021}. However, a notable research gap remains as these platforms show limitations in terms of their capacity to generate individualized UI components that are required for practical development projects~\cite{zhao_guigan_2021}.
    \item \textit{Enhancing usability and editability in design outputs:} We believe the primary requirement is the need for the system`s output to be in an editable format, allowing design modifications. Current systems focus mainly on UI image generation which require extra steps for editing~\cite{nandoskar_automated_2021}. If instead UI \textit{metadata} was generated in JSON format (for example), it would offer direct usability and further development~\cite{acornley_using_2021}. %The model`s training should prioritize generating modifiable UI components~\cite{bakaev_component-based_2018}, while enabling reusability and editability of components, which is yet to be introduced to the field. 
    This feature is yet to be introduced to the field. 
    
\end{itemize}

CoGen aims to address the above-mentioned gaps by means of the following:
\begin{itemize}
    \item Focusing on the fundamental building block of components (``Atoms''), rather than the final stage of UI designs. This is important because atoms are created in the initial stages of the design process and are used in subsequent stages. %each stage is a combination of the previous stage. Therefore, it is essential that atoms are created initially.
    \item Generation of JSON based on textual prompts (e.g., \textit{``Generate a Professional Button with a border radius of 10.0''}) to create reusable UI components. % that are compatible enough to be used with Figma.
    
\end{itemize}

Currently, Figma is one of the most popular tools that is used for UI design~\cite{staiano_designing_2022}. Therefore, we focus on creating UI components that are compatible enough to be used with Figma. Additionally, CoGen addresses the lack of a dataset which includes JSON-prompt pairs. We generate the requisite data pairs by training models on JSONs collected from publicly available Figma design sytems and from the \textit{Figcomponents} website~\cite{figcomponents}.

\section{Related Work}
Researchers have developed systems to assist the process of UI design, while addressing all stages of the design process~\cite{pitale_human_2019}. These projects range from basic support tools that provide output for design solutions~\cite{xu_ai_2023} and identification of design patterns to get UI design inspirations~\cite{nguyen_deep_2018}, to advanced systems capable of generating complete UI layouts~\cite{huang_creating_2021}.

\subsection{UI generation}
Some of the stages of UI design include design system creation, wireframing and prototyping.

\subsubsection{Wireframing}
Wireframes act as the blueprint of UI designs, aiding as an initial sketch of the intended UI design, where an overall layout is outlined. 

The authors of Akin developed a system that generates wireframes based on a selected design theme using a Self-Attention Generative Adversarial Network (SAGAN) model, which included a generator and dicriminator~\cite{gajjar_akin_2021}.

Sketch2Code is also a novel solution to detect hand-drawn UI wireframes such that they can be used to develop a UI representation object~\cite{jain_sketch2code_2019}.

\subsubsection{Lo-Fi and Hi-Fi prototyping}
Lo-Fi prototypes include more detailed aspects of a UI design, when compared to wireframes, while capturing the essential structural elements~\cite{xu_ai_2023}. In contrast, Hi-Fi prototypes are the finalised designs with colour, typography, structure, images, etc. 
An example of a system focusing on Hi-Fi designs is GUIGAN, which generates full UI design images. The authors utilise Sequential GANs, using a convolutional neural network (CNN) as the discriminator and a Long Short-Term Memory (LSTM) as the generator~\cite{zhao_guigan_2021}. By training the models on screenshots of existing applications, GUIGAN creates UI design images.

UIGenerator is another system designed for generating UI mockups, consisting of three key components. A \textit{UI generator} is used to generate the mockups using GANs. A \textit{multi-modal retriever} integrates text descriptions with UI data using Bidirectional Encoder Representations from Transformers (BERT). A \textit{text-only retriever} uses a text-based algorithm to retrieve UI designs from the training dataset that closely match the input description~\cite{huang_creating_2021}.

The authors of UIDiffuser use a stable diffusion model to generate UI designs for mobile devices. The system generates these designs from textual descriptions and predefined UI components~\cite{wei_boosting_2023}. 

Another notable project involves Large Language Models (LLMs) to generate UI layouts from textual descriptions, guided by a structured UI grammar. In this approach, models such as GPT-3 interpret the input descriptions, breaking them down into individual UI components. These components are then arranged according to predefined rules in the UI grammar and finally, the structured representation is displayed~\cite{lu_ui_2023}.

Analysing these projects, we observe that, while wireframing and prototyping have been the subject of significant research, the same cannot be said for design system creation. This is a gap that CoGen aims to fill.

\subsection{Prompt identification and generation}
%CoGen requires a component for prompt generation based on UI component JSON. Projects that closely align with these utilise Seq2Seq models or CNNs. 
%A number of projects, mostly utilising Seq2Seq models or CNNs, have focused on 

There are a number of projects related to prompt identification and generation. \textit{Prompt generation} is related to generating new textual prompts based on input data. \textit{Identification} indicates the ability to understand a given input in order to render an acceptable output.

The authors of UIGenerator use a text-only retriever, which relies on high-level text descriptions from the “screen2words” dataset~\cite{wang_screen2words_2021} to identify relevant UI designs. The method matches user-provided text descriptions with existing UI samples from the training set based on similarity~\cite{huang_creating_2021}.

Prompt Middleware involves a framework that generates prompts for LLMs based on the UI affordances. The study mentions three methods for generating prompts: \textit{static prompts} that are predefined to be mapped to the components, \textit{template-based prompts} that are partially predefined and include placeholders filled dynamically based on user interactions, and \textit{free form prompts} that are generated in real-time based on the user's input~\cite{macneil_prompt_2023}.

Widget Captioning focuses on natural language prompt generation for UI elements. The authors gathered data through crowd-sourcing. A Convolutional Neural Network (CNN) processes the visual input and extracts features, which are then passed to a Recurrent Neural Network (RNN) or transformer model to generate descriptions~\cite{li_auto_2020}. 

UI Grammar utilises LLMs like GPT-3 to generate mobile UI layouts from textual descriptions, guided by predefined rules specific to UI layouts. First, the input description is tokenised and processed according to these rules. The LLM interprets the context of the description and produces a layout in a structured format, such as JSON, which can be utilised by Figma~\cite{lu_ui_2023}.

Note that none of these projects generate textual outputs based on input UI JSON, which is necessary to complete the dataset required for CoGen (JSON-description pairs). 

% \subsection{Closely related projects}
% The Text2App project focuses on generating mobile applications from natural language (NL) specifications using Seq2Seq networks. The process begins by taking a description, which is then pre-processed using techniques like tokenization. The Seq2Seq model`s encoder processes the input text, converting it into a fixed-length context vector that captures the semantic meaning of the description. The decoder utilizes this context vector to produce an abstract intermediate formal language called Simplified App Representation (SAR). The SAR is then processed further to generate application code~\cite{hasan_text2app_2021}.

\subsection{Related Datasets}

% Requires: \usepackage{array}
\begin{table}[t]
    \centering
    \caption{Datasets for UI generation}
    \label{tab:ui-design-datasets}
    \begin{tabular}{|>{\raggedright\arraybackslash}p{0.2\linewidth}|>{\raggedright\arraybackslash}p{0.7\linewidth}|}
    \hline
    Rico & Rico~\cite{deka_rico_2017} is currently the largest existing dataset with designs from mobile applications. It holds data from more than 9,772 Android  applications under 27 categories. Of the 72,219 UI screens~\cite{zhao_guigan_2021} it holds, it includes the text, structure, visual and interactive design properties of the screens~\cite{gajjar_akin_2021}. Rico been used for UI related activities such as code generation, user prediction, layout related generation and other projects, and interaction modelling~\cite{sun_ui_2020}. \\
    \hline
    Screen2words & Screen2words includes UI screenshots and summaries that have been obtained via crowd-sourcing. The dataset is a bit lacking in terms of the language and the misalignment with UI terms~\cite{huang_creating_2021}.  \\
    \hline
    ERICA & ERICA~\cite{deka_erica_2016} consists of mobile app UI designs with user interactions, collected from apps in the Google Play store. ERICA includes 18,600 UI designs and 50,000 interactions approximately~\cite{malik_reimagining_2023}.  \\
    \hline
    REDRAW & REDRAW includes more than 190,000 UI components that are labelled in addition to more than 14,000 screens. They are collected from the Google Play store~\cite{malik_reimagining_2023}.  \\
    \hline
    \end{tabular}
    
\end{table}

Table~\ref{tab:ui-design-datasets} lists the most popular datasets used in ML-based UI design projects. 
In addition to the datasets listed in the table, there are also datasets for sketch-based inputs (e.g., UISketch~\cite{sermuga_pandian_uisketch_2021}), datasets for web designs that include font properties for text elements (e.g., CTXFonts Dataset~\cite{malik_reimagining_2023}) and more generic datasets for recognition of text-based information (e.g., ICDAR~\cite{karatzas_icdar_2015}). 

It is important to note that none of these datasets are closely relevant to CoGen. They lack JSON-prompt pairs that are essential to develop CoGen`s functionality.

\section{Approach and Methodology} \label{sec:approach}
Our approach consisted of two stages: The first stage involved prompt generation, while the second stage involved JSON generation using the previously generated prompts. These JSONs can then be used within Figma for component generation. We describe these stages further in Section~\ref{sec:main-methodology}. 

\begin{figure}
    \centering
    
    \frame{\includegraphics[width=\columnwidth]{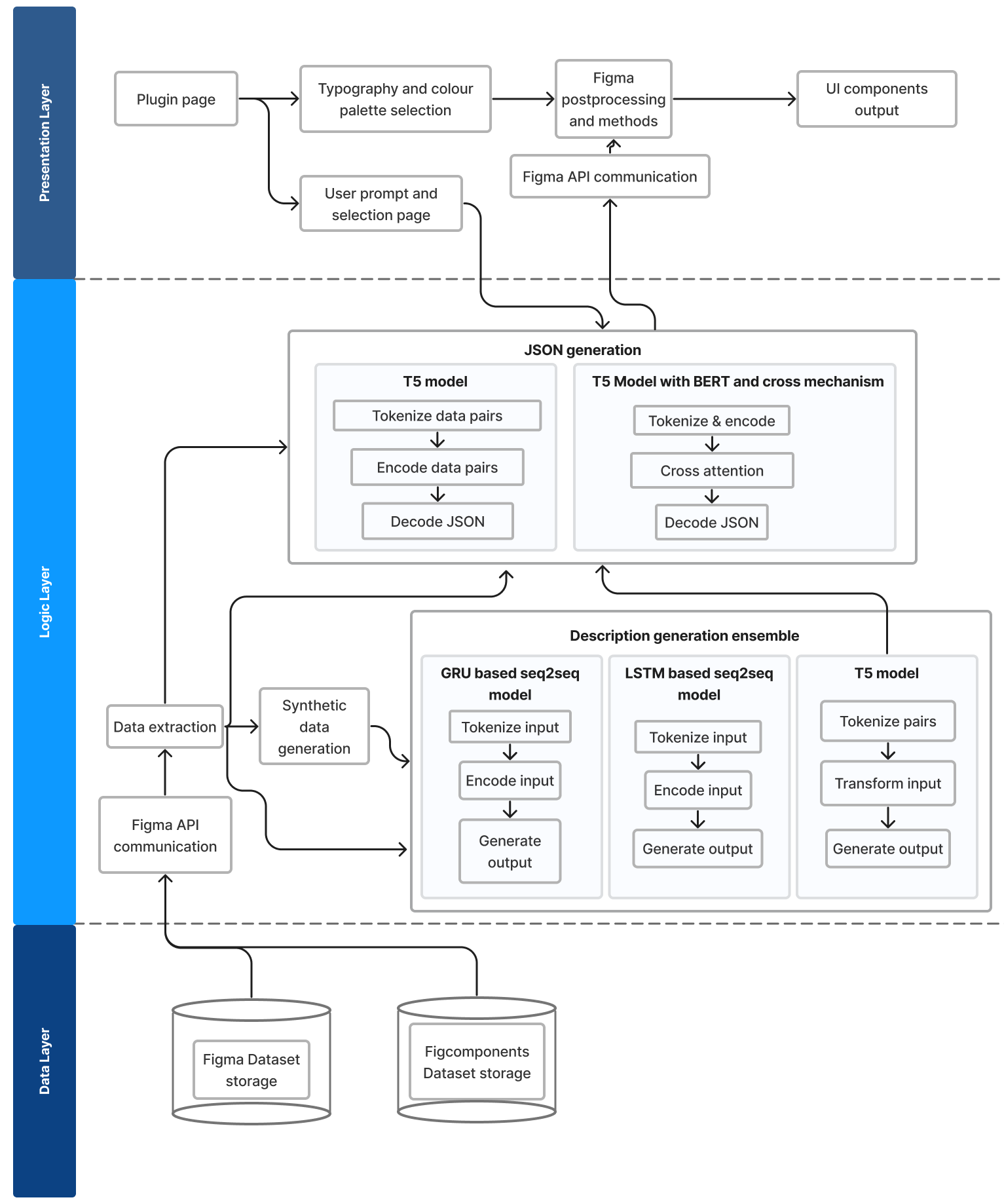}}
    \caption{Architecture Diagram}
    \label{fig:arc}
\end{figure}

Fig~\ref{fig:arc} shows the architecture diagram relevant to CoGen. 
The initial step involved the use of the Figma API to extract details about the UI components. Note that Figma allows designers to create components that include \textit{variants} within them. This is then known as a \textit{component set}. For example, a button component set could include a variant for the ``Default'' state and another variant for the ``Disabled'' state, each of which also has a ``Large'' and a ``Small'' variant. %This has been illustrated in Fig~\ref{fig:comp}. % TODO: recreate with correct sizes

% \begin{figure}
%     \centering
%     \includegraphics[width=1\linewidth]{figcomp.png}
%     \caption{Figma componentset and variant}
%     \label{fig:enter-label}
% \end{figure}

One important point to note then is the component naming style. The component sets use a predefined naming convention which splits the full component name as follows: 
\begin{align*}
    Style/ ComponentName/ Subtype
\end{align*}

\begin{itemize}
    \item Style - Explains the style or theme (\textit{`Basic'}, \textit{`Trendy'}, \textit{`Playful'}, \textit{`Professional'})
    \item ComponentName - The type of the component (\textit{`Button'}, \textit{`Input field'}, \textit{`Icon button'}, \textit{`Menu list'}, \textit{`List items'}, \textit{`Label'})
    \item Subtype - Additional details related to the components  (\textit{`Light'}, \textit{`Dark'}, \textit{`Default'}, etc).
\end{itemize}

% insert image

\subsection{Data extraction}
Given the unavailability of pre-existing datasets for the project, data was extracted from freely accessible Figma design systems and the Figcomponents website. The components were extracted in two different forms of JSON. The first version was a simplified flat structure, extracted without any nested hierarchy, while the second version was extracted by considering the nested hierarchies of the components. 

\begin{itemize}
    \item The simple JSON consists of properties like \textit{`colour`}, \textit{`stroke colour`}, \textit{`stroke weight`}, \textit{`text colour`}, \textit{`font family`}, \textit{`font weight`}, \textit{`font size`}, \textit{`effect name`}, \textit{`effect colour`}, \textit{`height`}, \textit{`width`} and \textit{`x`} and \textit{`y`} values. The variant details of a component will be extracted based on the naming. For instance, the naming convention for a variant in Figma is extracted in the form of a key-value pair. An excerpt has been shown in Fig~\ref{fig:json-from-figma}.
\end{itemize}

\begin{figure}
\begin{lstlisting}
{
    "variant_properties": {
        "color": "rgba(255, 255, 255, 1.0)",
        "strokes": [
            "rgba(126, 86, 216, 1.0)"
        ],
        "strokeWeight": 1.0,
        "text": "Button CTA",
        "textColor": "rgba(255, 255, 255, 1.0)",
        "borderRadius": 10.0,
        "fontFamily": "Inter",
        "fontWeight": 500,
        "fontSize": 14.0,
        "effects": [
            {
                "type": "DROP_SHADOW",
                "color": "rgba(16, 24, 40, 0.05000000074505806)"
            }
        ],
        "padding": 0,
        "width": 77.0,
        "height": 20.0,
        "x": -4619.0,
        "y": -2135.0,
        "hasIcon": false,
        "style": "Professional",
        "component_name": "Button",
        "subtype": "Default",
        "variant_details": {
            "State": [
                "Default"
            ],
            "Size": [
                "Small"
            ]
        }
    }
}
\end{lstlisting}
    \caption{Excerpt of JSON data from Figma}
    \label{fig:json-from-figma}
\end{figure}

\begin{itemize}
    \item The nested JSON is extracted based on hierarchical details and the properties are extracted based on the type of the node. For example. if the node type is `text`, font details will be extracted. If the node type is `vector', vector details such as height, colour and width will be extracted. There are also three ``grouping''-type nodes: \textit{frames}, \textit{autolayouts}, and \textit{groups}, where a node could contain one or more child nodes. If a node (parent or child) is of one of the grouping types, more generic details like the dimensions, colours, stroke details, effects, and border radius of the node will be extracted, along with information about its child nodes. This will continue recursively until no child nodes exist.
\end{itemize}

These two types of data (i.e., the simple and nested JSONs) were extracted in order to be used in the JSON generation stage (Section~\ref{sec:json-creation}) for comparing and evaluating the performance of different models.

\subsection{Prompt and JSON generation}\label{sec:main-methodology}
The major component of CoGen consists of two major parts: prompt creation based on JSON  input, and JSON creation based on the user input.

\subsubsection{Prompt generation} \label{sec:prompt-creation}
Initially we created synthetic textual prompts using the available JSON data from Figma. For example, if a field \texttt{border\_radius} exists for a component, one possible prompt would be \texttt{"Create a \{component\_name\} with border radius \{border\_radius\}."} This step was entirely rule-based. The data generated here was paired with the relevant JSON data from Figma and then fed into three models to compare and evaluate.

The models trained were an LSTM based Seq2Seq model, a Gated Recurrent Unit (GRU) based Seq2Seq model and a fine-tuned T5 model. 

\begin{figure}[t]
\begin{lstlisting}
sample_json = {
    "variant_properties": {
        "component_name": "Label", 
        "style": "Trendy", 
        "subtype": "Dark", 
        "variant_details": {
            "state": ["Enabled", "Focused"], 
            "size": ["Medium"]
        },
        "borderRadius": "8.0",
        "strokeweight": "2.0", 
        "effects": [
            {
                "type": "SHADOW", 
                "color": "rgba(0,0,0,0.2)"
            }
         ]
    }
}
\end{lstlisting}
    \caption{Sample JSON for T5}
    \label{fig:t5-sample-json}
\end{figure}

\begin{figure}[t]
\begin{lstlisting}
Generated Prompt 1: Generate a Trendy Label in any color color.
Generated Prompt 2: Create a Trendy Dark Label.
Generated Prompt 3: Build a Label with AA-Button styled Trendy with any color background and state of Supported, Size of Medium. 
Generated Prompt 4: A Label with border radius 88.0.
Generated Prompt 5: A Label with State of Active, Size of Medium.
\end{lstlisting}
    \caption{Sample prompts generated by T5}
    \label{fig:t5-sample-description}
\end{figure}

For the LSTM and GRU based models, a BERT tokenizer was used to process text into tokens, while a T5 tokenizer was applied for the T5 model. The JSON-prompt pairs were extracted and saved as tensors for the LSTM and GRU models. 

The LSTM-based model included an embedding layer for the input tokens, and LSTM layers for the encoder and decoder. There was also a Fully Connected (FC) layer to produce the output tokens. When training, an Adam optimizer was used with a learning rate of 0.001 and a step learning rate scheduler to adjust the learning rate for every epoch. Moreover, the model implemented gradient clipping to prevent exploding gradients and validated the model at each epoch, saving the best model based on validation loss.

The GRU-based model included an embedding layer for the input tokens, and GRU layers for the encoder and decoder. There was also a Fully Connected (FC) layer to generate final token predictions. In the dataset we used, the ‘button’ class was overrepresented. To address this imbalance in the dataset, while training, the GRU-based model included class weights in the loss function to handle imbalanced datasets for more accurate training.

While training the T5 model, the data was converted into HuggingFace’s~\cite{huggingface} dataset format for compatibility with the pre-trained encoder-decoder architecture. This processed input sequences and generated corresponding output sequences specific to this project to generate prompts. Here, a ‘t5-small’ model was used. %The training arguments configure training settings including the output directory, number of epochs, batch size, gradient accumulation, warmup steps, weight decay, logging, evaluation strategy, mixed precision training, and checkpoint saving.

The results from the three models were evaluated (see details in Section~\ref{sec:eval-desc-gen}) and the T5 model was found to outperform the other two. Figures~\ref{fig:t5-sample-json}
 and~\ref{fig:t5-sample-description} show the input and output, respectively, from the fine-tuned T5 model.
 
\subsubsection{JSON creation}\label{sec:json-creation}
The JSON creation stage involves providing a prompt that is processed to output a JSON that can be used with Figma. 

Two fine-tuned T5 models were compared to create the output. The first model was a simple fine-tuned T5 while the second model included cross-mechanism with BERT embeddings. 

For the simple fine-tuned T5, the data is first loaded and tokenized. Each prompt is converted to token IDs and sequences are padded to the longest sequence in the batch and truncated if they exceed a predefined maximum length. Then the JSON is converted to a string and tokenized. Additionally, the process  includes T5's own attention mask, which is a binary mask indicating which tokens are actual input tokens (1) and which are padding (0)~\cite{wang_codet5_2021}. 

For the complex T5 model, the BERT tokenizer converts prompts to token IDs and generates BERT embeddings. Then the T5 tokenizer is used to convert both prompt and JSON to token IDs. 
This model is extended to include a cross-attention mechanism with BERT embeddings. It initialises a multi-head attention layer for cross-attention and a linear layer to project BERT embeddings to the T5 hidden state dimension. In the forward method, input IDs are passed through the T5 encoder to obtain hidden states, BERT embeddings are projected and expanded to match the sequence length, and cross-attention is applied between T5 hidden states and BERT embeddings. The cross-attention hidden states are then passed through the T5 decoder to generate the output sequence, and the loss is computed if labels are provided, handling any shape mismatches between logits and labels.

\subsection{Post-processing}
Before generating the components in Figma, post-processing plays a essential role in ensuring the accuracy and usability of the JSON output as obtained from the JSON creation stage (Section~\ref{sec:json-creation}). The post-processing stage involves validating the JSON data to check for syntax errors and invalid characters. Once validated, the JSON is mapped to a format compatible with Figma, following the nested structure for representing the components.

Following this step, each component type is processed using specifically customised methods to generate the corresponding nodes. For instance, a button component may include a frame node containing a text node inside it. Similarly, for icons, a default rectangle is used as a placeholder to maintain structural consistency.

% \begin{figure}[ht]
% 	\centering
% 	\includegraphics[width=0.45\textwidth]{Fig2}
% 	\caption{Red, blue and green line}
% 	\label{fig2}
% \end{figure}

\section{Results and Evaluation}
We discuss the evaluation results for the two stages relevant to the development of CoGen below.

\subsection{Prompt generation}\label{sec:eval-desc-gen}
%The most important part in prompt generation is the identification of the component name. 
Since the prompt generation stage focuses on the creation of prompts that completely align with an input JSON, it is important that the accuracy in terms of the component name is high. To calculate the accuracy metrics, subsets were used for 100,200,300,400,500 samples. This was done to evaluate the model's performance in identifying component names across varying sample sizes.
The process involved loading the dataset into a dictionary format, tokenizing inputs with the T5 tokenizer, and progressively increasing the sample size to generate prompts from the model, which were then compared with the exact values from the JSON input. The accuracy metrics relevant to this process have been presented in Table~\ref{tab:acc_tex}.

\begin{table}[t]
    \centering
    \caption{Prompt generation accuracy metrics}
    \label{tab:acc_tex}
    
    \begin{tabular}{|c|c|c|c|c|}

        \hline
        \textbf{Model} & \textbf{Accuracy\%} & \textbf{F1\%} & \textbf{Recall\%} & \textbf{Precision\%} \\\hline
        LSTM based & 56.8 & 52.58 & 56.8 & 55.18 \\ \hline
        GRU based & 96.2 & 96.08  & 96.2 & 96.52  \\ \hline
        T5 & 98.0 & 98.2 & 98.0  & 98.44 \\ \hline
    \end{tabular}

\end{table}

 %Therefore, from the accuracy metrics, we considered the T5 model to be the best option. 
We also performed a second assessment to evaluate the variety in the generated prompts, i.e., how much the sentence structure of one prompt differed from another while retaining the same meaning. For this, we computed BLEU scores and ROUGE scores for the three models. These are shown in Table~\ref{tab:scr_tex}.

\begin{table}[t]
    \centering
    \caption{Prompt generation scores}
    \label{tab:scr_tex}
    
    \begin{tabular}{|c|c|c|c|c|}

        \hline
        \textbf{Model} & \textbf{BLEU} & \textbf{ROUGE-1 } & \textbf{ROUGE-2 } & \textbf{ROUGE-L } \\\hline
        LSTM based & 0.2515  & 0.8758  & 0.7657  & 0.8752 \\ \hline
        GRU based & 0.2402 & 0.9738   & 0.9581 & 0.9734  \\ \hline
        T5 &  0.2668 & 0.5486  & 0.3967   & 0.5293  \\ \hline
    \end{tabular}
    
\end{table}

We note that a BLUE score of 0.2668 indicates that the T5 model is learning to generate diverse prompts that are different from the dataset, while its accuracy score (98\%+) shows that it extracts the component name accurately in the prompts. More importantly, the T5 model also worked well with unseen JSON and created prompts that reflected that JSON. We therefore opted to use the T5 model for prompt generation.

% \begin{figure}
%     \centering
%     \includegraphics[width=1\linewidth]{images/Sample JSON for T5.png}
%     \caption{Sample JSON for T5}
%     \label{fig:enter-label}
% \end{figure}

\subsection{JSON generation}
JSON generation was evaluated with two types of data: simple JSON and nested JSON. The BLEU and ROUGE metrics were calculated for 100 prompt samples taken from the generated prompt dataset (Section~\ref{sec:prompt-creation}). The results are shown in Table~\ref{tab:json_simple} and Table~\ref{tab:json_nested}.
\begin{table}[t]
    \centering
    \caption{Metrics for Simple JSON}
    \label{tab:json_simple}
    
    \begin{tabular}{|c|c|c|c|c|}

        \hline
        \textbf{Model} & \textbf{BLEU} & \textbf{ROUGE-1 } & \textbf{ROUGE-L } \\\hline
        Simple T5  & 0.6071  & 0.6659  & 0.6433  \\ \hline
        T5 with x-attention and BERT  & 0.5574 & 0.6142   & 0.5940  \\ \hline
        
    \end{tabular}
    
\end{table}

\begin{table}[t]
    \centering
    \caption{Metrics for Nested JSON}
    \label{tab:json_nested}
    
    \begin{tabular}{|c|c|c|c|c|}

        \hline
        \textbf{Model} & \textbf{BLEU} & \textbf{ROUGE-1 } & \textbf{ROUGE-L } \\\hline
        Simple T5  & 0.3769   & 0.4544  & 0.2414  \\ \hline
        T5 with x-attention and BERT  & 0.3771  & 0.4511    & 0.2503  \\ \hline
        
    \end{tabular}
    
\end{table}

\begin{table}[t]
    \centering
    \caption{Accuracy metrics for Simple JSON}
    \label{tab:acc_simple}
    
    \begin{tabular}{|c|c|c|c|c|}

        \hline
        \textbf{Model} & \textbf{Accuracy} & \textbf{F1} & \textbf{Recall} & 
\textbf{Precision } \\\hline
        Simple T5  & 86.2\%  & 89.6\% & 86.2\% & 94.2\%   \\ \hline
        T5 with x-attention and BERT & 97.4\%  & 97.6\%  & 97.4\% & 98.2\%   \\ \hline
        
    \end{tabular}
    
\end{table}

\begin{table}[t]
    \centering
    \caption{Accuracy metrics for Nested JSON}
    \label{tab:acc_nested}
    
    \begin{tabular}{|c|c|c|c|c|}

        \hline
        \textbf{Model} & \textbf{Accuracy} & \textbf{F1} & \textbf{Recall} & 
\textbf{Precision} \\\hline
        Simple T5  & 100\%  & 100\%  & 100\% & 100\%   \\ \hline
        T5 with x-attention and BERT & 98\%  & 99\%  & 98\% & 98.5\%  \\ \hline
        
    \end{tabular}
    
\end{table}

Based on the results, we see that the simple T5 model produces the best results for simple JSON, whereas for the nested JSON, the complex T5 model works better. 

Table~\ref{tab:acc_simple} and Table~\ref{tab:acc_nested} show the accuracy, F1 score, precision and recall related to the component type identification. 

Further testing was done to evaluate the results obtained from the models, in order to test their functionality in terms of actual prompts. Five prompts encompassing the different component types, styles and other properties were used, and evaluation was based on the following criteria:

\begin{itemize}
    \item \textbf{Variant details and properties identified:} For example, the size of a button being small in size/drop shadow effects/ border radius effects, stroke weights etc.
    \item \textbf{ComponentName identification: }For example, if the prompt says \textit{‘Create a button’} then the \textit{`ComponentName}` would be \textit{‘button’}.
    \item \textbf{Style identification:} For example, if the prompt says \textit{‘Create a Professional button’} then the style should be equal to \textit{‘Professional’}.
    \item \textbf{JSON keys reflect those in the dataset:} All the keys present in the dataset should be reflected in the output regardless of syntax errors in the JSON.
\end{itemize}

Each of the factors is evaluated with a binary output of Pass/ Fail, and a weightage of 25\% each for each prompt. Table~\ref{tab:simple_success_rates_simplet5} and Table~\ref{tab:simple_success_rates_complext5} provide an overview of the results for the simple T5 model and the T5 model with BERT and cross-mechanism, respectively, for simple JSON inputs. Tables~\ref{tab:nested_success_rates_simplet5} and~\ref{tab:nested_success_rates_complext5} provide the same results for nested JSON inputs.

% Requires: \usepackage{graphicx}
\begin{table}[t]
    \centering
    \caption{Success Rates for Different Prompts for Simple JSON with T5 fine-tuned model}
    \label{tab:simple_success_rates_simplet5}
    \begin{tabular}{|l|c|c|c|}
    \hline
    \textbf{Prompt} & \textbf{Pass} & \textbf{Fail} & \textbf{Success rate} \\ \hline
    Button & 5 & 0 & 100\% \\ \hline
    Label & 5 & 0 & 100\% \\ \hline
    Input fields & 3 & 2 & 60\% \\ \hline
    Menu & 3.75 & 1.25 & 75\% \\ \hline
    List-item & 5 & 0 & 100\% \\ \hline
    Icon button & 3.75 & 1.25 & 75\% \\ \hline
    \end{tabular}
    
\end{table}
% Requires: \usepackage{graphicx}
\begin{table}[t]
    \centering
    \caption{Success Rates for simple JSON with Different Prompts with T5 with Cross mechanism and BERT embeddings}
    \label{tab:simple_success_rates_complext5}
    \begin{tabular}{|l|c|c|c|}
        \hline
        \textbf{Prompt} & \textbf{Pass} & \textbf{Fail} & \textbf{Success rate} \\
        \hline
        Button & 3 & 2 & 60\% \\ \hline
        Label & 4.5 & 0.5 & 90\% \\ \hline
        Input fields & 2 & 3 & 40\% \\ \hline
        Menu & 1.75 & 3.25 & 35\% \\ \hline
        List-item & 3.25 & 1.75 & 65\% \\ \hline
        Icon button & 4.75 & 0.25 & 95\% \\ \hline
        
    \end{tabular}
    
\end{table}

\begin{table}[t]
    \centering
    \caption{Success Rates for Different Prompts for Simple JSON with T5 fine-tuned model}
    \label{tab:nested_success_rates_simplet5}
    \begin{tabular}{|l|c|c|c|}
    \hline
    \textbf{Prompt} & \textbf{Pass} & \textbf{Fail} & \textbf{Success rate} \\ \hline
    Button & 3.5 & 1.5 & 70\% \\ \hline
    Label & 3 & 2 & 60\% \\ \hline
    Input fields & 3.5 & 1.5 & 70\% \\ \hline
    Menu & 3.25 & 1.75 & 65\% \\ \hline
    List-item & 2.5 & 2.5 & 50\% \\ \hline
    Icon button & 3 & 2 & 60\% \\ \hline
    \end{tabular}
    
\end{table}
\begin{table}[t]
    \centering
    \caption{Success Rates for Different Prompts with T5 with Cross mechanism and BERT embeddings}
    \label{tab:nested_success_rates_complext5}
    \begin{tabular}{|l|c|c|c|}
        \hline
        \textbf{Prompt} & \textbf{Pass} & \textbf{Fail} & \textbf{Success rate} \\
        \hline
        Button & 4.5 & 0.5 & 90\% \\ \hline 
        Label & 4 & 1 & 80\% \\  \hline
        Input fields & 2.75 & 2.25 & 55\% \\ \hline
        Menu & 3.75 & 1.25 & 75\% \\ \hline
        List-item & 3 & 2 & 60\% \\  \hline
        Icon button & 3 & 2 & 60\% \\
        \hline
    \end{tabular}
    
\end{table}

Based on the results, the simple JSON data when trained with the simple T5 model produced the best results. Therefore it is evident that manipulating the simple JSON is a better option for Figma integration. 

% Another reason for favouring the simple JSON was that the complex v failed to identify inner details in the nested JSON despite having the correct JSON structure. Therefore, the model chosen was the simple T5 fine-tuned model. 

\begin{figure}
    \centering
    \includegraphics[width=1\linewidth]{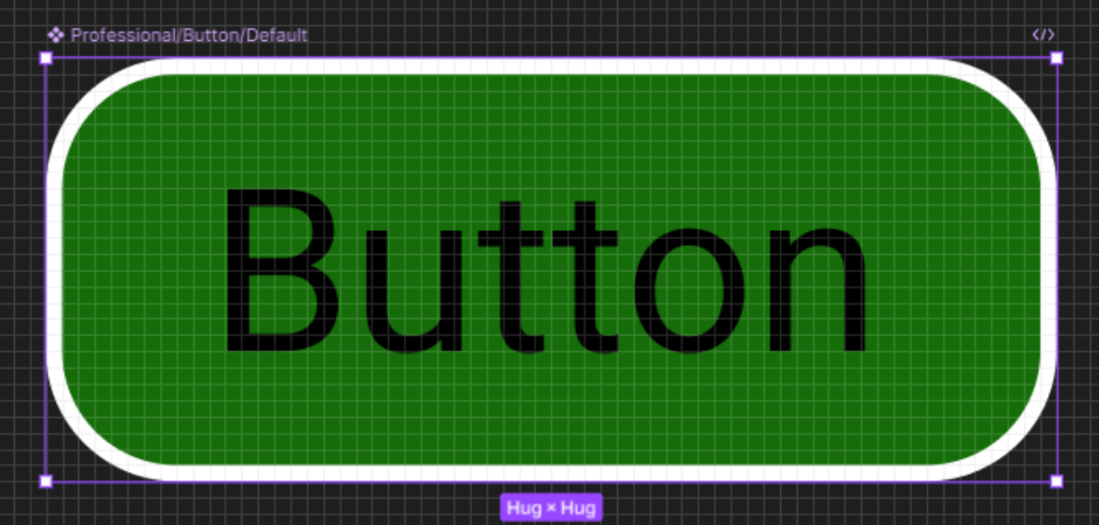}
    \caption{Figma output}
    \label{fig:out}
\end{figure}

Fig~\ref{fig:out} shows the output from Figma for the prompt \textit{`generate a professional button with a size of small`} when using the simple T5 model with simple JSON input.

\section{Limitations and Future Work}
We discuss here the limitations of CoGen and possible avenues for future work.

\begin{itemize}
    \item \textbf{Limited number of components:} CoGen only works with six components (\textit{’Button’}, \textit{’Input field’}, \textit{’Icon button’}, \textit{’Menu list’}, \textit{’List items’}, \textit{’Label’}), and four styles (\textit{`Basic`}, \textit{`Trendy`}, \textit{`Playful`}, \textit{`Professional`}). This is mainly due to resource limitations and the lack of publicly available Figma datasets. If more datasets or Figma design systems are made available, the project could be expanded to have more basic components and also nested components that use combinations of these basic components. 

    \item \textbf{Creation of component sets with variants:} As mentioned in Section~\ref{sec:approach}, Figma allows for the creation of component sets with multiple variants of a component within. However, at present, CoGen is able to create only a single component which includes one variant with a single prompt. This again is mostly due to time and resource constraints relating to the training of models. In future, CoGen could be expanded to include more variants within a component set, with more advanced prompting techniques and usage of models, which could then enable developers with no prior knowledge of Figma to create fully-fledged UI designs.

    \item \textbf{Lack of a vast dataset: }The T5 model requires more input with nested components in order to understand the relationship between the node type (frame, text, icon) and the properties of each node. This will result in better component generation within Figma as well. 

    \item \textbf{Lack of complex models:} The model’s ability to understand diverse prompts is very limited. Due to the lack of resources we could not access more diverse pre-trained models to test with. Access to paid models like GPT-4 or GPT-5 would make this process much faster and more efficient as they are more refined in terms of understanding prompts and generating content. Moreover, this would aid in creating nested components as well.

    \item \textbf{Lack of complex integrations:} The ability to integrate with sites such as Google Fonts to include more font options, or connecting to an icon library set, will be a huge added advantage to ensure an efficient design process.
    
\end{itemize}

\section{Conclusion}
In this paper, we have presented CoGen, an ML-based tool for generating Figma-compatible UI components from user-provided text prompts. %uses ML techniques by comparing several models. This was useful in contributing towards the domain and therefore can be considered a contribution towards similar systems where prompts are analysed and used to create structured data like JSON.
The project presents a novel approach to UI component generation, addressing a gap in current research. Unlike existing systems that primarily focus on the overall visual design or image-based representations, this work emphasizes the creation of UI components based on input prompts.
Additionally, prompt generation from component JSON adds significant value to the UI design field, as most datasets typically consist only of design details, rather than the descriptive information needed to guide component creation.

\bibliographystyle{IEEEtran}

\bibliography{bibliography}

\end{document}